\documentclass[aps,prb,superscriptaddress,reprint,showpacs]{revtex4-1}

\usepackage{color}
\usepackage{amsmath}
\usepackage{amssymb}
\usepackage{graphicx}
\usepackage[squaren]{SIunits}

\begin{document}

\title{Bloch band dynamics of a Josephson junction in an inductive environment}

\author{T. Wei\ss l}
\email{thomas.weissl@grenoble.cnrs.fr}
\affiliation{Institut N{\'e}el, CNRS et Universit{\'e} Joseph Fourier, BP 166, F-38042 Grenoble Cedex 9, France}
\author{G. Rastelli}
\affiliation{Universit{\"a}t Konstanz, Fachbereich Physik, 78457 Konstanz, Germany}
\affiliation{Universit{\'e} Grenoble 1/CNRS, LPMMC UMR 5493, B.P. 166, 38042 Grenoble, France}
\author{I. Matei}
\affiliation{Institut N{\'e}el, CNRS et Universit{\'e} Joseph Fourier, BP 166, F-38042 Grenoble Cedex 9, France}
\author{I. M. Pop}
\affiliation{Department of Applied Physics, Yale University, New Haven, Connecticut 06520, USA}
\affiliation{Institut N{\'e}el, CNRS et Universit{\'e} Joseph Fourier, BP 166, F-38042 Grenoble Cedex 9, France}
\author{O. Buisson}
\affiliation{Institut N{\'e}el, CNRS et Universit{\'e} Joseph Fourier, BP 166, F-38042 Grenoble Cedex 9, France}
\author{F. W. J. Hekking}
\affiliation{Universit{\'e} Grenoble 1/CNRS, LPMMC UMR 5493, B.P. 166, 38042 Grenoble, France}
\author{W. Guichard}
\affiliation{Institut N{\'e}el, CNRS et Universit{\'e} Joseph Fourier, BP 166, F-38042 Grenoble Cedex 9, France}
\begin{abstract}
We have measured the current-voltage characteristics of a Josephson junction with tunable Josephson energy $E_J$ embedded in an inductive environment provided by a chain of SQUIDs. Such an environment induces localization of the charge on the junction, which results in an enhancement of the zero-bias resistance of the circuit. We explain this result quantitatively in terms of the Bloch band dynamics of the localized charge. This dynamics is governed by charge diffusion in the lowest Bloch band of the Josephson junction as well as by Landau-Zener transitions out of the lowest band into the higher bands.
In addition, the frequencies corresponding to the self-resonant modes of the SQUID array exceed the Josephson energy $E_J$ of the tunable junction, which results in a renormalization of $E_J$, and, as a consequence, an increase of the effective bandwidth of the lowest Bloch band.
\end{abstract}


\date{\today}

\maketitle

\section{Introduction}
\label{intro} Superconductors provide the unique possibility to create dissipationless macroscopic electrical quantum circuits that
are characterized by the dynamics of a well-defined single degree of freedom, the superconducting phase.
This is due to the macroscopic coherence of the superconducting wave function~\cite{Tinkham}. In Josephson junction (JJ) circuits, the phase difference $\hat{\varphi}$ between the two superconductors forming the junction and $\hat{N}$ the number of Cooper pairs that tunnel through the junction are quantum conjugate variables satisfying $\left[ \hat{\varphi},\hat{N}\right] = i$. The relative strength of the fluctuations is proportional to the square root of the ratio of the charging energy $E_C$ and the Josephson energy $E_J$, $\Delta\varphi / \Delta N \sim \sqrt{E_C/E_J}$.
Therefore, the Josephson effect in circuits containing
large Josephson junctions ($E_J\gg E_C$) enables the control of a well-defined dissipationless phase state. This feature allowed the realization of the metrological Volt standard using networks of JJs~\cite{Kautz1996}.\\
Decreasing the junction size into the submicron region decreases the ratio $E_J/E_C$ and, as a consequence the quantum fluctuations of the phase, $\Delta{\varphi}$, start to play a dominant role. Large quantum phase fluctuations can produce $\Delta{\varphi}\approx 2\pi$ windings, a phenomenon known as quantum phase-slips. Since increasing phase fluctuations implies decreasing fluctuations of its conjugate variable, the charge, the resulting well-defined charge state is expected to yield insulating behavior of the junction. However, such a state is not easily observed for a single Josephson junction, which is typically measured in a superconducting low-impedance environment. The associated charge relaxation time $\tau_q$ is too short to preserve the well-defined charge state on the measurement time scales.\\
The relevance of Josephson junction circuits with a well-defined charge state has been pointed out in pioneering articles by Averin, Likharev and Zorin~\cite{Likharev1985,Averin1985}. Indeed, Josephson junctions with a well-defined charge should sustain Bloch oscillations, consisting of voltage oscillations\cite{Nguyen2007} on the junction due to a periodic motion of the charge in the lowest Bloch band. These Bloch oscillations are dual to the standard Josephson oscillations. Therefore, they could be used in principle in quantum metrology to realize a current standard, analogous to the way Josephson oscillations are used to realize a voltage standard.\\
A possible strategy to obtain long charge relaxation times consists of integrating a highly resistive element with resistance $R$ near the junction in order to increase the charge relaxation time $\tau_q= RC$ enabling the realization of a Coulomb blockade state. However, the introduction of a dissipative element in the circuit introduces heating effects along with Johnson-Nyquist noise making it difficult to reach the quantum limit of minimum charge fluctuations~\cite{Webster2013}.\\
The first experiments used single Josephson junctions with on-chip, highly resistive environments~\cite{Kuzmin1991,Likharev1985} in order to observe Bloch oscillations. Later, Josephson junction chains in the resistive state have been used to create an environment resistive enough to observe the so-called Bloch nose~\cite{Watanabe2001,Corlevi2006a}. More recently, longer chains have been studied~\cite{Ergul2013} and the zero-bias resistance has been interpreted in terms of quantum phase-slips. In the limit of dominating charging energy, the zero bias resistance can be understood in terms of depinning of charges in the chain~\cite{Vogt2014}.\\
Alternatively, nanowires have been suggested as superconducting elements sustaining phase-slips. Due to the low-dimensionality of these wires, phase-slips occur easily~\cite{Arutyunov2008}, thereby reducing the fluctuations of the charge. Inspired by Ref.~\onlinecite{Mooij2006}, several experiments were performed using superconducting nanowires probing the dual physics of these systems such as coherent quantum phase-slips in a device dual to the Cooper pair box~\cite{Astafiev2012}, a quantum phase-slip interference device dual to a Superconducting Quantum Interference Device (SQUID)~\cite{Hongisto2012,Pop2012}, and attempts to measure Bloch oscillations and dual Shapiro steps~\cite{Lehtinen2012,Webster2013}.\\
Another possible strategy to obtain long charge relaxation times is the additional integration of a large inductance close to the junction ~\cite{Guichard2010,Manucharyan2009,Masluk2012,Gershenson2012}.  Indeed, in the resulting $RLC$ circuit, the charge relaxation time will be dominated by $\tau_q=L/R$, as soon as $L/R \gg RC$. Hence the use of a highly inductive environment enables one to achieve long charge relaxation times with small resistance $R$, thus avoiding the aforementioned heating effects. Experiments performed in the group of M. Devoret at Yale, probing the quantum states of the so-called fluxonium qubit, demonstrated that charge fluctuations on a small junction can be reduced by embedding it in a highly inductive environment~\cite{Manucharyan2009,Pop2014}.\\
In this article we present an experimental study and a quantitative analysis of the zero bias resistance induced by Bloch band dynamics of the charge on a tunable Josephson junction included in an inductive environment. The Bloch bands result from coherent quantum phase-slips occuring on the tunable junctions. The dynamics of the system is related with the presence of the series inductance that plays the role of an effective mass. We demonstrate that this configuration allows the observation of charge localization.\\
From the measurements we infer that the dynamics of the charge is a combination of thermal hopping in the lowest Bloch band and Landau-Zener processes causing interband transitions. In addition, we account for the effects of the zero point motion of the electromagnetic modes in the JJ chain and show that they result in a renormalization of the Josephson coupling energy $E_J$, and hence in an increase of the bandwidth of the lowest Bloch band of the tunable junction.
\section{Single junction in an inductive environment}
\label{singjunc}
\begin{figure}
\includegraphics[width=0.4\textwidth]{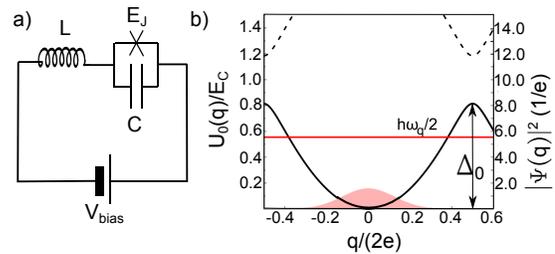}
\caption{a) Circuit diagram of a voltage-biased Josephson junction in an inductive environment. b) Wave-function localized in the effective potential $U_{0}\left(q \right)$ of the lowest Bloch band (black continuous line). The second band is indicated by the black dashed line. The red horizontal line corresponds to the lowest energy level. The graph corresponds to the parameters used in the experiment with a flux frustration of f=0.494.}
\label{fig:1}
\end{figure}
Before presenting and discussing the experimental results, it is useful to briefly recall the theory of a single Josephson junction in an inductive environment \cite{Likharev1985,Guichard2010}. The circuit is presented in Fig.~\ref{fig:1}a. It contains a junction with a capacitance $C$, such that the charging energy is given by $E_C = e^2/2C$, and a Josephson energy $E_J$. The junction is coupled to a series inductance $L$.
The single junction is described by the standard Hamiltonian
\begin{equation}
H = \frac{Q^2}{2C} - E_J \cos \varphi.
\label{ham}
\end{equation}
The eigenfunctions are Bloch states and the energy spectrum is described by Bloch bands as a function of the corresponding quasi-charge $q$. If the ratio $E_J/E_C$ is small, the bands are almost parabolic with a width
\begin{equation}
 \Delta_0 \sim E_C
\label{eq:Delta0}
\end{equation}
and with gaps $E_{gap}\sim E_J$. If $E_J/E_C$ is large, the bands are sinusoidal with a width $\Delta_0 \approx (E_J^3 E_C)^{1/4} e^{- \sqrt{8E_J/E_C}}$ and with gaps $E_{gap}\sim \hbar \omega_p$, where $\omega_p = \sqrt{8E_J E_C}/\hbar$ is the junction's plasma frequency.
As long as LZ transitions between the bands can be ignored, the behavior of the junction is completely governed by the properties of the lowest band.  Outside the parameter range $E_J \gg E_C$, the bands are not sinusoidal and the dependence of the lowest band $U_0$ on the quasi-charge $q$ is given by
\begin{equation}
U_{0}\left(q\right) =\sum_{k=1}^{\infty} \nu_k \cos{\left(k\pi q/e\right)} \,\label{eq:Ham_qps_2}.
\end{equation}
The non-sinusoidal nature of the bands is reflected by the summation over higher harmonics with index $k$, which corresponds to the quantum phase slip processes in the junction where the phase winds by an amount $2\pi k$. The energies $\nu_k$ are the amplitudes for these multiple windings to occur.
By embedding the Josephson junction  in an inductive environment, it is possible to induce dynamics of the quasi-charge $q$ in the lowest band. The characteristic kinetic energy is then given by $E_{L}={[\Phi_{0}/(2\pi)]}^{2}/2L$, where $\Phi_0$ is the superconducting flux quantum. The full Hamiltonian for the corresponding circuit, shown in Fig.~\ref{fig:1}a for a voltage-biased configuration, is given by~\cite{Mooij2005,Mooij2006,Guichard2010}
\begin{equation}
H=-\left(\frac{\hbar^{2}}{2L}\right)\frac{\partial^{2}}{\partial q^{2}}+U_{0}\left(q \right)-V_{bias} q\,.\label{eq:Ham_qps_3}
\end{equation}
The inductance plays the role of the mass of a fictitious particle with coordinate $q$, moving within the potential energy $U_0(q)$\cite{Guichard2010}. We denote $\Delta_0$ the barrier height separating the minima of the potential $U_0(q)$.
In the tight-binding limit, $E_L\ll \Delta_0$, at vanishing voltage bias $V_{bias}=0$, we can use the harmonic approximation for the potential $U_{0}\left(q\right)\simeq q^{2}/(2C_{q})$
where the effective capacitance is defined as $C_{q}^{-1}={(\partial^{2}U_{0}/\partial q^{2})}_{q=0}$~\cite{Likharev1985}. In this case, the ground-state wave function is
a Gaussian whose width equals to $\Delta q^{2}=(e^{2}/\pi)R_{Q}/\sqrt{L/C_{q}}$
with $R_{Q}=h/(4e^{2})\simeq$ $6.45$ k$\Omega$. The ground-state energy equals $\hbar\omega_{q}/2$,
where 
\begin{equation}
\omega_{q}=1/\sqrt{LC_{q}}
\label{eq:wq}
\end{equation}
is the associated dual plasma frequency.\\
In our experiment we can change the ratio $E_{J}/E_{C}$ of the
Josephson junction {\em in situ} for a given inductance $L$, thus changing the barrier height $\Delta_{0}$, and explore in particular the regime $\Delta_{0}>\hbar\omega_{q}/2$  where a localized charge-state is expected to appear, see Fig.~\ref{fig:1}b. We have measured the zero-bias resistance as a function of the ratio $E_{J}/E_{C}$ and analyze the result in terms of a possible charge localization.
\section{System}
\label{system}
\begin{figure}
\includegraphics[width=0.5\textwidth]{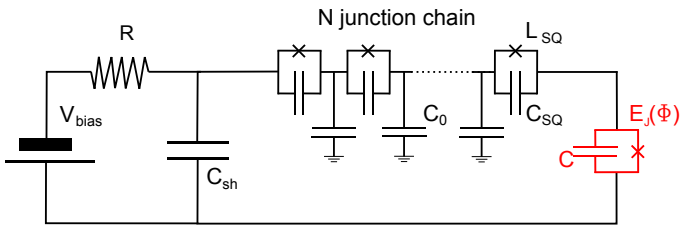}
\caption{Experimental circuit showing the voltage-biased tunable Josephson junction (printed in red) connected to a Josephson junction chain with $N$ junctions. Each junction is a SQUID loop. The tunable junction and the chain are protected from spurious high-frequency noise by the parallel arrangement of the shunt capacitance $C_{sh}=\unit{200}{pF}$ and the resistance $R=\unit{3}{k\Omega}$.} \label{fig:2}
\end{figure}
The experiments were performed on the circuit shown in Fig.~\ref{fig:2}. In order to realize a Josephson junction with tunable $\Delta_0$, we designed it in the form of a SQUID such that the ratio $E_J/E_C \propto |\cos( \pi \Phi/\Phi_0)|$, where $\Phi$ is the magnetic flux threading the SQUID loop. For future use we define the flux frustration parameter $f = \Phi/\Phi_0$.\\
The $Al/AlO_x/Al$ SQUIDs are fabricated using two-angle shados evaporation of aluminum on a silicon substrate with \unit{100}{nm} of silicon oxide. The two aluminum layers have a thickness of $\unit{20}{nm}$ and $\unit{40}{nm}$. The junction area inferred from SEM-images is $\unit{0.07}{\mu m^2}$ and the lattice parameter of the SQUID chain is $\unit{350}{nm}$.\\
The junction capacitance is estimated from the design to be $C=\unit{6.9}{fF}$ such that $E_{C}/h=\unit{2.8}{GHz}$. At $f=0$ the Josephson energy of the tunable junction is $E_{J}/h=\unit{29}{GHz}$.\\
The series inductance consists of a linear chain of $N$ SQUIDs, each characterized by a Josephson inductance $L_{SQ}=\unit{10}{nH}$ for an applied magnetic field that corresponds to $f=0.5$ for the tunable junction. The inductance $L_{SQ}$ was estimated from resistance measurements at $T=\unit{1.5}{K}$. This chain provides therefore a total inductance of $L= N L_{SQ}$. This inductance is also flux-dependent. The SQUID loops forming the chain are 1.6 times smaller than the SQUID loop forming the tunable junction. The inductance of the chain $L$, changes only by $\unit{10}{\%}$ in the small flux frustration range $f=[0.46,0.5]$ where the tunable junction is probed, so that in first order it can be considered as constant.  We fabricated chains containing a different number $N$ of junctions $N=28, 38, 48, 68, 88$ and $108$ \footnote{The measured samples comprise two sets of four chains (8, 18, 28, 38) and  (48, 68, 88, 108). The two sets have slightly different parameters. As the chains with 8 and 18 squids do not show charge localization we did not include them in the discussion. The chains are actually measured in a configuration where one chain is connected in series with the parallel combination of the other three chains on the chip. We corrected the zero bias resistance for the presence of the other three chains.}. The capacitance of the SQUIDs in the chain is equal to the one forming the tunable junction. The capacitance to ground of the islands between the SQUIDs is estimated to be $C_{0}\simeq C_{SQ}/75$. The sample is shunted by a home-made $NbTi/Al_{2}O_{3}/NbTi$ parallel capacitor $C_{sh}=\unit{200}{pF}$. This shunt capacitor, together with a resistance $R =\unit{3}{k\Omega}$,  provides a low-pass filter with a cut-off frequency of about 60 MHz, thus protecting the junction and its inductive environment from spurious noise above this frequency. The measurements were carried out in a two-point configuration using high-frequency filters in form of thermocoax cables. We measure the I-V characteristics of the array comprising the single junction and the inductive Josephson chain, as depicted in Fig.\ref{fig:2}. In the small region at $f=[0.46,0.5]$, where we will analyze our data, the circuit is effectively voltage biased. The sample is kept at the base temperature of about \unit{50}{mK} of our dilution refrigerator\cite{Balestro}.
\section{Experimental results and qualitative discussion}
\label{qualitative}
\begin{figure}[!h]
\includegraphics[width=0.5\textwidth]{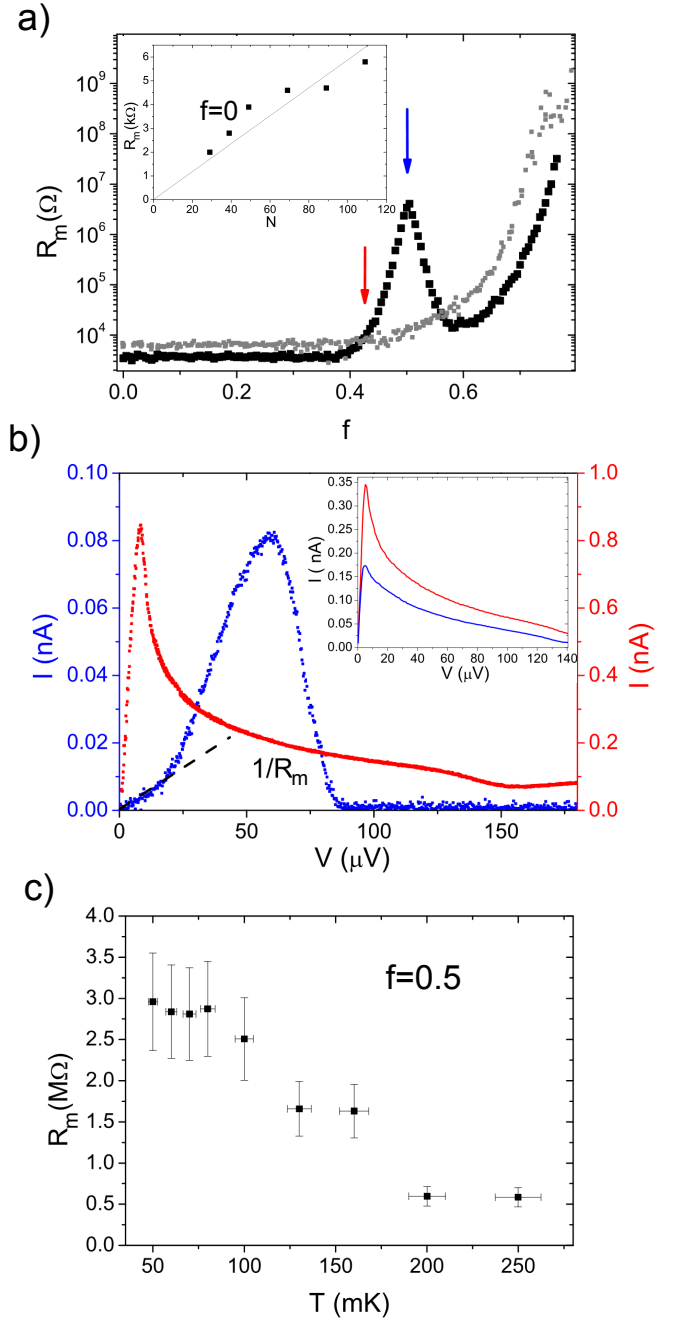}
\caption{a) Measured zero-bias resistance as a function of flux frustration parameter $f$ for a tunable junction connected to a 48-junction Josephson junction chain (black). The gray curve corresponds to a Josephson junction chain with 49 junctions but without the tunable junction. The inset shows the dependance of the zero-bias resistance of chains with tunable junctions as a function of N for $f=0$. b) Current-voltage characteristics of the tunable junction taken at $f=0.42$ and $f=0.5$. These flux values are marked in a) by the two arrows. The inset shows the corresponding I-V characteristics for the uniform chain. c) Zero-bias resistance as a function of temperature for $f=0.5$.  }
\label{fig:3}
\end{figure}

We start our discussion by focusing on the results obtained for the single junction in series with the 48 junction chain as an example to illustrate the results. Figure \ref{fig:3}a shows the zero-bias resistance of this circuit as a function of the flux frustration parameter $f$.\\
First of all we note that the resistance is a few k$\Omega$ for zero frustration. We attribute this finite resistance value to the occurrence of residual incoherent quantum phase-slips in the junctions forming the chain. We can estimate the rate\cite{Caldeira1983} for a phase winding by $\pm 2\pi$ for a given junction as $\Gamma_{qps}^{\pm} = A e^{-B^{\pm}}$, where 
\begin{equation}
A= 12 \sqrt{6 \pi}\frac{\omega_p}{2\pi} \sqrt{\frac{\Delta U}{\hbar \omega_p}} \mbox{, } B^{\pm} = \frac{36}{5}\frac{\Delta U^{\pm}}{\hbar \omega_p}
\end{equation}
and where we neglected the change of the barrier height due to the current bias in the prefactor.
Here, $\Delta U \sim 2 E_J-\hbar\omega_p/2$ and $\Delta U^{\pm} \sim 2 E_J-\hbar\omega_p/2\pm \pi I \hbar/2e $ is the effective barrier for tunneling of the phase of the junction by $\mp 2 \pi$. Since the temperature is lower than the plasma frequency $\omega_p/2\pi=\unit{25.4}{GHz}$, thermal activation can be ignored. Quantum phase-slips give rise to a voltage 
\begin{equation}
V=h (\Gamma_{qps}^+-\Gamma_{qps}^{-})/2e.
\label{eq:volt}
\end{equation}
Linearizing (\ref{eq:volt}) with respect to the bias current $I$ we find $V = R_{qps} I$ with the following estimate for the zero-bias resistance:
\begin{equation}
R_{qps} \simeq R_Q \frac{36\pi}{5} \frac{\Gamma_{qps}}{\omega_p},
\end{equation}
where $\Gamma_{qps}= 2 A \exp{(\frac{36}{5}\frac{\Delta U}{\hbar \omega_p})}$.
Using the parameters of the experiment, we find a residual resistance per junction of $R_{qps}\approx\unit{10}{\Omega}$. \\ 
This zero-bias resistance at small values of the frustration changes linearly with the number $N$ of the junctions (see inset of Fig ~\ref{fig:3}a). From the experimental N dependence we extract a resistance of $\unit{59}{\Omega}$ per junction which is of the same order of magnitude as the theoretical estimate.\\ %For low frustration of the chain the resistance shows no flux dependence. This indicates that the bottle-neck for the incoherent phase-slips is actually not the tunnel rate but rather the relaxation rate.
The Josephson junction chain is a highly refractive material so that photons can only propagate slowly. This arises the question whether or not the processes ocurring on the single junction are influenced by all the junctions in the Josephson junction chain or not. We therefore estimate the horizon, that is the distance which photons can propagate within the timescales of phase and charge tunneling processes on the single junction. The relevant time scales on which this processes happen are given by $\Delta t_\varphi = \hbar/(2eV_{bias})$ for the phase tunneling and $\Delta t_q = 2e/ I $ for charge tunneling\cite{Devoret1990}. We use $V_{bias} =\unit{20}{\mu V}$ and $I=\unit{10}{pA}$ (see figure \ref{fig:3} b) which yields $\Delta t_\varphi=\unit{16}{ps}$ and $\Delta t_q = \unit{32}{ns}$. \\
The phase velocity (in units of junctions per second) can be estimated as $v_p= \omega / k = N/ (2\pi \sqrt{L_{SQ} C_0}) \approx 10^{13}$ junctions/s. This results in a horizon for the phase tunneling of $\approx 160$ junctions and $\approx 3\ 10^5$ junctions for the charge tunneling so that in both cases the entire Josephson junction chain contributes to the electromagnetic environment of the junction.
\\
Upon increasing the frustration $f$ on the tunable junction, we observe a significant increase of the zero-bias resistance $R_m$ up to 4 M$\Omega$ reached for the maximal frustration $f=0.5$ (black curve in figure \ref{fig:3} a). As the SQUIDs in the Josephson junction chain have smaller surface their resistance remains almost unchanged. For comparison we plot the zero bias resistance of a uniform Josephson junction chain (gray curve in figure \ref{fig:3}a). In this work we concentrate on the resistance peak around $f=0.5$ of the larger SQUID. This flux bias corresponds to a situation where phase-slips predominantly occur on the single junction and the juctions chains act like an inductance. The second peak in the zero bias resistance at higher flux frustraton corresponds to the situation where phase-slips occur on the SQUIDs in the junction chain. This regime will not be discussed in the paper. Figure \ref{fig:3}b shows the current-voltage characteristics, taken at the two flux frustrations corresponding to the onset of the resistance increase ($f=0.42$, left arrow in Fig.~\ref{fig:3}a) and at its maximum ($f=0.5$, right arrow in Fig.~\ref{fig:3}a). The inset show the I-V characteristics of the uniform chain at the same flux biases. We understand this resistance increase by three orders of magnitude as a result of the enhanced charge localization within the first Bloch band. As the flux frustration $f$ is increased, the ratio $E_J/E_C$ of the junction decreases, thereby increasing the width $\Delta_0$ of the lowest Bloch band. As soon as the bandwidth $\Delta_0$ exceeds the kinetic energy $E_L$ due to the inductance, the charge becomes more and more localized. More precisely, in this limit the dual plasma frequency becomes $\omega_q /2 < \Delta_0/\hbar$ and localized states can form in the minima of the the lowest band $U_0(q)$.
We estimate the barrier height of the potential $U_0(q)$ to be $\Delta_0/h \approx 2.4$ GHz and the dual plasma frequency to be $\omega_q /2\pi=\unit{4}{GHz}$. An estimation of the ratio $\Delta_{0}/E_{L}=0.6$
for our quantum phase-slip junction results in a situation where only
one level, with energy $\hbar\omega_{q}/2$, is located in the potential
$U_{0}(q)$, see Fig.~\ref{fig:1}b.\\
At higher voltages, the I-V characteristic for $f=0.5$ shows a current peak that we attribute to the existence of electromagnetic modes in the chain at higher frequencies\cite{Hofheinz2011,Ingold}. The effect of these modes on the behavior found at low bias will be discussed in detail in Sec.~\ref{effcirc}. For voltage biases much larger than the plasma frequency of the JJ chain $V_{bias}\gg \hbar \omega_{SQ} /2e$ photons can no longer be emitted to the environment so that the incoherent charge tunneling is suppressed and a zero current state is observed. \\
In Fig.~\ref{fig:3}c, the temperature dependence of the zero-bias resistance is shown for $f=0.5$. We see that with decreasing temperature $T$, the zero bias resistance increases, down to a temperature of about 80 mK. At lower temperatures, $R_0$ saturates. The effect of finite temperatures is to induce thermal fluctuations kicking the charge particle out of the minima. The gradual decrease of $R_0$ with increasing $T$ is therefore expected: charge localization is more pronounced at lower temperatures.  The low-temperature saturation is most probably a result of the existence of a finite environment noise-temperature of about 80 mK. We estimated that quantum tunneling can still be ignored at this temperature. Similar findings were reported in Ref. ~\onlinecite{Webster2013}.

\section{Charge diffusion}
\label{chargediff}
\begin{figure}
\includegraphics[width=0.35\textwidth]{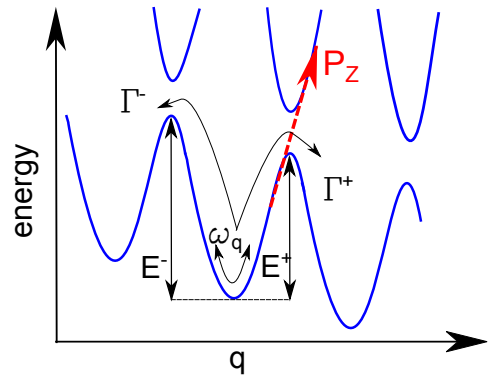}
\caption{Escape rates $\Gamma^+$ and $\Gamma^-$ for the fictitious quasi-charge particle moving in a tilted washboard potential. $P_Z$ denotes the Landau-Zener probability for the quasi-charge particle to undergo an interband transition.} \label{fig:4}
\end{figure}

In order to account for the dependence of the measured zero-bias resistance on the flux frustration parameter $f$, we start by analyzing the simplest possible model describing the dynamics of the quasi-charge $q$, assuming its dynamics to be restricted to the lowest charge band. This amounts to ignoring Landau-Zener transitions.  We will come back to this assumption in section~\ref{zener} below. For now we consider the limit of low voltage bias and low temperature, $eV, k_BT < \Delta_0$. We include the effect of a small tilt due to finite bias, see Fig.~\ref{fig:4}. Classically, the charge will be localized in one of the minima of the band. This would give rise to a zero current state. However, driven by thermal or quantum fluctuations, the particle can hop between neighboring minima. In the presence of a finite bias voltage, the rate $\Gamma^-$ for hopping from right to left differs from $\Gamma^+$ corresponding to hopping from left to right, giving rise to a finite current $I = 2e (\Gamma^+ - \Gamma^-)$.
We calculate the rates $\Gamma^\pm$ using Kramers's classical result\cite{Kramers1940} for the escape of a particle from a potential well.
We assume thermal activation to be dominant  as the temperature is in the same orders of magnitude as the dual plasma frequency $\omega_q /2\pi=\unit{4}{GHz}$ so that the rates $\Gamma^\pm$ can be expressed as
\begin{equation}
\Gamma^\pm = \frac{\omega_q}{2 \pi}e^{- E^\pm/k_BT}.
\end{equation}
Here we used the attempt frequency $\omega_A=\omega_q $, and $E^\pm = {\Delta_0} \mp eV$ denote the barrier heights for tunneling to the left and the right of the well in the presence of the tilt $V$ (see also Fig.~\ref{fig:4}). We recall that the parameter ${\Delta_0}$ depends on the ratio $E_J/E_C$ of the tunable junction that is varied {\em in situ} through the flux frustration parameter $f$.\\ The current flowing through the system is $I = V/R_0$, and linearizing the rates with respect to a small bias voltage $V$ results in
\begin{equation}
R_0 = R_Q\frac{k_B T}{\hbar \omega_q}e^{{\Delta_0}/k_BT}.
\label{R0fit}
\end{equation}
\begin{figure}
\includegraphics[angle=0,width=0.5\textwidth]{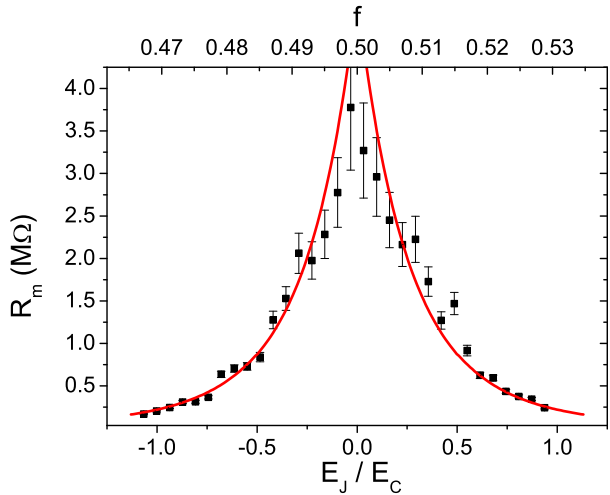}
\caption{Zero-bias resistance $R_0$ as a function of the parameter $E_J/E_C$ for the chain with $N=48$: experimental results (data points) and fit (solid line) based on Eq.~(\ref{R0fit})} \label{fig:5}
\end{figure}
\begin{table}
\center
\begin{tabular}{c|c|c|c}
N & $\alpha$ & $\omega_q^{fit} / E_C $ &$ \omega_q  / E_C $\\
\hline
\hline
28 &$1.7 \pm 0.1$ & $0.61 \pm 0.02$& 1.17 \\
\hline
38 &$4.1 \pm 0.1$ & $7.0 \pm 0.4$& 1.01\\
\hline
48 &$4.6 \pm 0.2$ & $1.7 \pm 0.5$&  0.99 \\
\hline
68 &$5.4 \pm 0.2$ & $17 \pm 4$& 0.84 \\
\hline
88 &$6.3 \pm 0.2$ & $52\pm 12$&0.74\\
\hline
108 &$5.5 \pm 0.2$ & $22 \pm 7$&0.67
\end{tabular}
\caption{Fitting parameters and errors for all measured chains. $\alpha$ is the multiplicative factor in front of $\Delta_0$ used when fitting with Eq.~(\ref{R0fit}).}
\label{tab:1}
\end{table}
In Fig.~\ref{fig:5}, the measured zero-bias resistance $R_m$ is shown as a function of frustration and the corresponding ratio $E_J/E_C$ for the $N=48$ junction chain, together with a fit based on Eq.~(\ref{R0fit}). Similar fits have been performed for all the measured chains. These fits enable us to compare the behavior of the fitted exponent $\Delta_0^{fit}$ as a function of the ratio $E_J/E_C$ with the theoretically expected one $\Delta_0$, determined by solving the Mathieu equation corresponding to Hamiltonian (\ref{eq:Ham_qps_3}) using the system's parameters. From the fits we find that the actual values of the exponent and the prefactor differ greatly from the expected one. Specifically, we find that good fits can be obtained using a barrier height $\Delta_0^{fit}=\alpha \Delta_0$ that is $\alpha$ times higher than the expected value $\Delta_0$. In Table~\ref{tab:1}, we list the multiplicative factors $\alpha$ for the other samples together with the prefactor $\omega_q^{fit}$ obtained by fitting and the corresponding prefactor expected from theory $\omega_q$. Note that the factor $\alpha$ is an almost monotonically increasing function of the chain length $N$. We obtain values ranging from $\alpha = 1.7$ to $\alpha = 6.3$. 
As will be argued in the following, we can attribute this discrepancy between the experimentally found exponent and the theoretically expected one to two phenomena: (i) the renormalization of the Josephson coupling energy of the tunable junction due to electromagnetic modes propagating along the chain and (ii) the effect of interband transitions (Landau-Zener processes) that dominate the charge dynamics whenever the gap $\sim E_J$ separating the lowest two charge bands becomes too small compared to the characteristic energy of the dynamics of the quasi-charge. This will happen when the flux bias on the single junction approaches $f=0.5$.
\section{Renormalization of the bandwidth}
\label{effcirc}
\begin{figure}
\includegraphics[angle=0,width=0.5\textwidth]{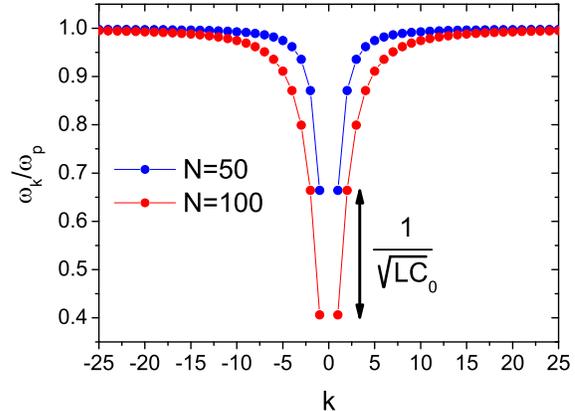}
\caption{Dispersion of the propagating modes on Josephson junction chains with a finite capacitance to ground $C_0/C_{SQ} = 1/50$.} \label{fig:6}
\end{figure}
It is well-known that the Josephson coupling energy $E_J$ of a Josephson junction connected to an external circuit is suppressed down to a value $E_J^*$ by the quantum fluctuations induced by the corresponding environment~\cite{Schmid1983, Schoen1990, Weiss2008}. As a consequence, the bandwidth $\Delta_0^*$ found for the lowest charge band will be larger than the bare width $\Delta_0$. Referring again to the measurement circuit as shown in Fig.~\ref{fig:2}, we first note that the junction is well-protected from high-frequency noise originating from the external leads by thermo coax and the RC filter down to 60 MHz\cite{Balestro}. Since the resistance measurements were performed at flux frustration parameters $f$ corresponding to a bare Josephson coupling energy in the GHz range, we suppose that the external noise does not account for any renormalization of $E_J$. However, the chain providing the inductance is directly coupled to the tunable junction. The fact that the islands realize a finite capacitance $C_0$ to ground leads to the appearance of dispersive electromagnetic modes in the chain~\cite{Rastelli2013,Masluk2012,Pop2011}, with a dispersion relation given by
$$
\omega_k = \omega_p \sqrt{\frac{1 - \cos k}{1-\cos k +C_0/2C_{SQ}}},
$$
where $k = 2 \pi n/N$ are the discrete wave-vectors of these modes. Here we assumed periodic boundary conditions as we do not know the exact boundary conditions in our experiment. The dispersion relation is shown in Fig.~\ref{fig:6}. It consists of a linear part, $\omega_k = \omega_0 k $ for small $k$-vectors, with $k < \sqrt{C_0/C_{SQ}}$.  Here $\omega_0 = 1/\sqrt{L C_0}$. The frequency $\omega_l$ of the lowest mode is inversely proportional to the chain length $N$. For the longest chains measured, this frequency is estimated to be about 40 $\%$ of the chain's plasma frequency $\omega_p$. For larger wave-vectors, the dispersion relation saturates at the plasma frequency $\omega_p$. In units of temperature, the frequency range between $\omega_l$ and $\omega_p$ covered by the modes corresponds to a range between $\unit{300}{mK}$ and $\unit{1}{K}$. Since the temperature at which the experiment is performed is lower than $\unit{300}{mK}$, a zero-temperature treatment of the modes is adequate. The equivalent voltage range is between $\unit{30}{\mu V}$ and $\unit{100}{\mu V}$, which corresponds to the range where the current peak is found in the I-V characteristic, see Fig.~\ref{fig:3}b, thereby providing indirect evidence for the existence of the modes. In this relatively high bias voltage range, Cooper pair transfer in the small junction is possible as the modes of the chain provide the necessary dissipation.\\
\begin{figure}
\includegraphics[angle=00,width=0.5\textwidth]{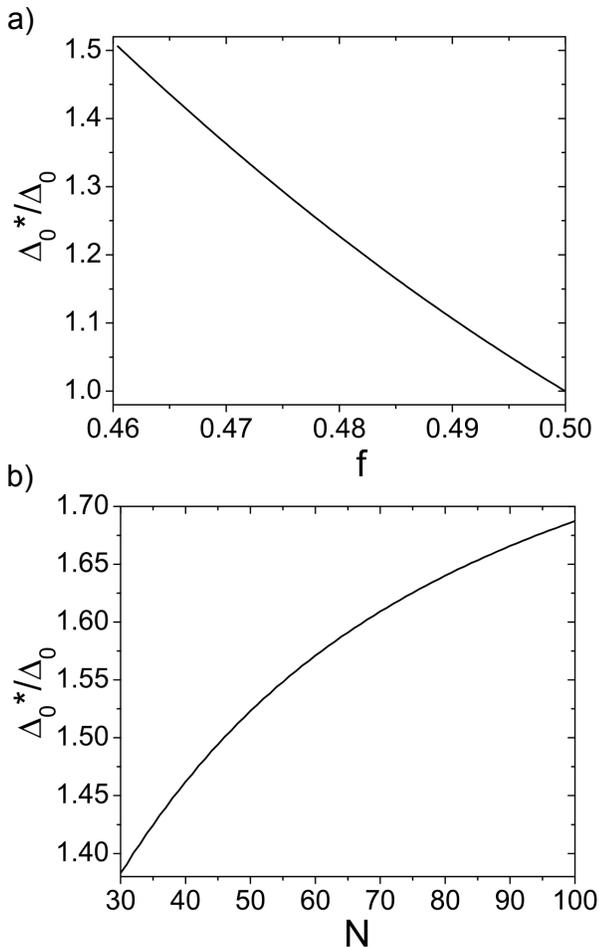}
\caption{Dependence of the ratio $\Delta_0^*/\Delta_0$ of the tunable junction's renormalized bandwidth and the bare bandwidth as a function of a) frustration for a chain length of $N=48$, b) chain length for a frustration $f=0.46$.  } \label{fig:7}
\end{figure}
In the limit of small voltage biases, as it is argued in Appendix~\ref{harmodes}, the modes induce zero point quantum phase fluctuations at the end of the chain that couple to the small junction. They add to the phase difference $\varphi$ across the junction, thereby renormalizing the Josephson coupling energy $E_J$ of the tunable junction \cite{Hekking1997} down to a value $E_J^*$, given by
\begin{equation}
E_J^* = E_J \exp\left\{-\frac{1}{N}\sum _k \frac{e^2}{\hbar}\sqrt{\frac{L}{2C(k)[1-\cos k]}} \right\}.
\label{ejstar}
\end{equation}
Using this result, we calculate the renormalized effective bandwidth $\Delta_0^*$, corresponding to the spectrum of Hamiltonian (\ref{ham}). Figure \ref{fig:7}a shows the dependence of the ratio $\Delta_0^*/\Delta_0$ as a function of $f$, for the $N= 48$ junction chain. We see that this ratio is larger than unity and a monotonically decreasing function of $f$, reaching unity at $f= 1/2$, where the Josephson coupling energy vanishes and the bandwidth attains its largest possible value $E_C$. The $N$-dependence of the ratio $\Delta_0^*/\Delta_0$ is plotted in Fig.~\ref{fig:7}b, for a fixed value of $f= 0.46$. Indeed,
it is a monotonically increasing function of $N$, however, even for $N=108$ it does not exceed a value of about $1.7$, and therefore can account only in part for the observed discrepancy discussed in section~\ref{chargediff} above. In the next Section we will show that Landau-Zener processes between the lowest charge band and the higher ones may account for the rest of the discrepancy.
\section{Charge diffusion in the presence of Landau-Zener processes}
\label{zener}
In this section we extend the charge diffusion model presented in Sec.~\ref{chargediff} and include the effect of possible interband transitions ignored so far. The probability to pass from the lowest charge band to the next one is given by~\cite{Schoen1990}
\begin{equation}
P_Z = \exp \left\{- \frac{\pi ^2} {4} \frac{E_{gap}^2} {E_C \hbar \omega_x}\right\},
\label{PZ}
\end{equation}
where $E_{gap}\sim E_J$ is the gap separating the two bands and $\omega_x$ the relevant frequency associated with the dynamics of the quasi-charge. At zero voltage bias, this frequency will be approximately given by the attempt frequency $\omega_q$ at which the quasi-charge tries to escape by thermal activation from the well formed by the minima of the lowest band. In presence of dissipation, the Landau-Zener probability is determined by coupling of the quasi-charge with the external enviroment. This can give rise to an effective gap appearing in Eq.(\ref{PZ}). \cite{Ingold,Schoen1990}
We observe that Landau-Zener processes are flux-dependent. We take into account the dissipative corrections by allowing $\omega_x$  in Eq.(12) to be an independent fit parameter. The probability to remain in the lowest band is given by $1-P_Z$. We then expect the measured resistance $R_m$ to be given by the weighted sum
\begin{equation}
R_m = (1-P_Z) R_0 + P_Z R_Z,
\label{fitform}
\end{equation}
where $R_0$, the resistance associated with charge diffusion in the lowest band (Eq.~(\ref{R0fit})) is now calculated with a multiplicative factor $\alpha=1$ but taking into account the renormalized barrier height $\Delta_0^*$. $R_Z$  is a constant fit parameter, representing the resistance associated with charge dynamics in the higher bands.
\begin{figure}
\includegraphics[angle=0,width=0.5\textwidth]{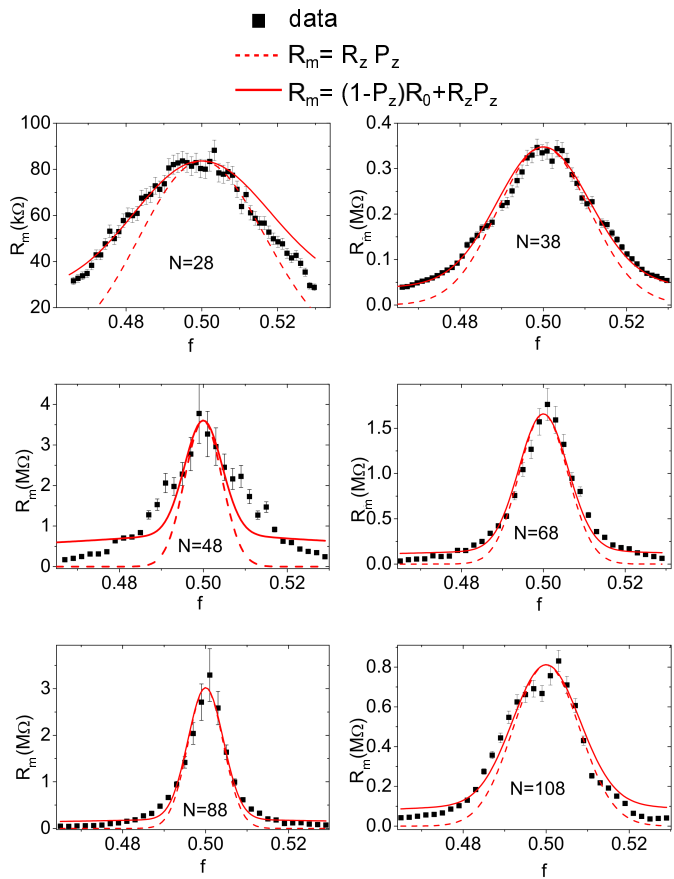}
\caption{Zero-bias resistance $R_0$ as a function of the parameter f for all the chains discussed in this work: experimental results (data points), fit (solid line) based on Eq.~(\ref{fitform}). The dashed line shows the contribution of Landau-Zener interband transitions alone $ P_Z R_Z$. } \label{fig:8}
\end{figure}
We assume $R_Z$ to be independent of the flux frustration parameter. When calculating $E_{gap}$ in Eq.~(\ref{PZ}), we solve the Mathieu equation associated with Hamiltonian (\ref{ham}) for the single junction, using the renormalized Josephson energy $E_J^*$, Eq.~(\ref{ejstar}). In Fig. \ref{fig:8} we show a fit of the data for all our chains with Eq.~(\ref{fitform}). 
Table ~\ref{tab:2} shows the parameters used to obtain the best fits.\\
\begin{table}
\center
\begin{tabular}{c|c|c|c}
N &  $R_{Z}\, (k\Omega)$& $\omega_x / E_C $ & $\omega_q^{fit} / E_C $ \\
\hline
\hline
28 & $84 \pm 2$  & $0.046 \pm 0.001 $& $0.535 \pm 0.005$\\
\hline
38 & $350 \pm 10$  & $0.020 \pm 0.001 $& $0.400 \pm 0.010$\\
\hline
48 & $3600 \pm 200$  & $0.005 \pm 0.002 $& $0.025 \pm 0.005$\\
\hline
68 & $170 \pm 100 $  & $0.010 \pm 0.003 $& $0.120 \pm 0.010$\\
\hline
88 & $3000 \pm 200$  & $0.005 \pm 0.001 $& $0.100 \pm 0.020$\\
\hline
108  & $800  \pm 50$  & $0.017 \pm 0.002 $& $0.170 \pm 0.030$
\end{tabular}
\caption{Fitting parameters and errors for all measured chains for fits using Eq.~(\ref{fitform}).}
\label{tab:2}
\end{table}
A few remarks are in order at this point. Note that $\omega_q^{fit}$ has the tendency to decrease with the chain length, a fact that is expected as the attempt frequency of the escaping particle decreases with increasing particle mass, the mass of the charge being given by the chain's inductance. On the other hand, the frequency $\omega_q^{fit}$ and the fitted LZ frequency $\omega_x$, are systematically smaller than the frequency $\omega_q$ (see Table \ref{tab:1} and \ref{tab:2}) associated with the curvature of the lowest Bloch band. This indicates that the charge motion is possibly overdamped \cite{Wubs2006}. Such overdamped motion could result from a finite quality factor of the electromagnetic modes. Indeed, microwave transmission experiments\cite{Weissl} done on a 200 Josephson junction chain with a similar ratio $E_J/E_C\approx10 $ as in the experiment here have shown an internal quality factor of about 100. We note that the fitting parameter $R_Z$ increases as a function of $N$ and takes a maximum value of $R_Z=\unit{3500}{\Omega}$ for $N=48$. Above $N=48$ it is difficult to conclude a systematic behavior of $R_Z$.  A more detailed understanding of the behavior of $R_Z$  requires a more detailed understanding of the processes responsible for charge relaxation in the higher bands, which is beyond the scope of the present paper.\\
In summary, as it can be seen in Fig. \ref{fig:8}, the measured zero-bias resistance as a function of the flux frustration parameter shows three different behaviours. Close to $f=0.5$, where $E_J \ll E_C$, Landau-Zener processes dominate. The measured resistance has a peak, the form of which is entirely dominated by the flux-dependence of the Landau-Zener probability $P_Z$. Away from $f=0.5$, the decrease of the resistance with $f$ becomes slower than the one predicted by $P_Z$ alone: the system enters the regime where charge diffusion within the lowest band dominates. Even farther away from $f=0.5$, the charge diffusion model breaks down. Here, the bandwidth becomes smaller than the residual noise temperature of the system that we estimate to be about $80$ mK. This happens at $f \approx 0.47$. The charge is no longer localized and our model based on simple escape rates underestimates the actual charge transfer rates. Hence the fit over-estimates the measured resistance. 
\section{Conclusions}
Our transport measurements suggest the existence of a localized charge state on a Josephson junction due to an inductive environment. We could explain the measured zero-bias resistance with a model combining charge diffusion within the lowest Bloch band and Landau-Zener processes between bands. In order to reduce the charge dynamics to the lowest Bloch band, future experiments should study single Josephson junctions with a larger Josephson coupling $E_J$ hence avoiding Landau-Zener transitions. At the same time the inductance of the environment should be increased to ensure a localized charge state. Increasing the SQUID chain length to obtain a larger inductance reduces the frequencies of the electromagnetic modes and renormalisation effects will start to play a dominant role. Future experiments using the combination of a tunable junction with a controllable SQUID chain constitute an experimental test bed to explore the renormalization effects of the electromagnetic environment on a small junction in a more systematic way. The understanding of the interplay between the charge dynamics and the electromagnetic environment is also relevant for future applications where an inductive environment could play an important role such as current Shapiro steps in a superconducting phase-slip circuit\cite{Guichard2010,Webster2013} or a quantum phase-slip qubit\cite{Mooij2005}.
\section{acknowledgements}
T.W. acknowledges support from the Grenoble Nanoscience Foundation.
G.R acknowledges  support from the EU FP7 Marie Curie Zukunftskolleg Incoming Fellowship Programme,
University of Konstanz (grant no. 291784). F.H. and W.G. are supported by Institut universitaire de France. W.G. also acknowledges support from the European Research council (grant no. 306731). The authors thank N. Roch for fruitful discussions and M. Vanevi{\'c} for useful comments on the manuscript.
\appendix
\section{Harmonic modes of a Josephson junction chain}
\label{harmodes}
\begin{figure}
\includegraphics[width=0.5\textwidth]{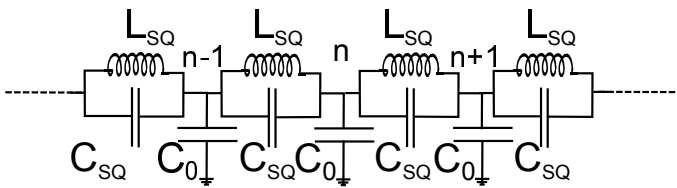}
\caption{Josephson junction chain.} \label{fig:A}
\end{figure}
In this Appendix we briefly review the quantum theory of a harmonic Josephson junction chain. We consider a Josephson junction chain, consisting of $N$ junctions, each with a capacitance $C_{SQ}$ and a Josephson coupling energy $E_{J,ch}$. We denote the capacitance of the islands between the junctions to ground by $C_0$. In the harmonic limit, valid when $E_{J,ch} \gg E_C = e^2/2 C$,  the junctions forming the chain behave as inductances with inductance $L_{SQ} = (\hbar/2e)^2/E_{J,ch}$ each. Then the chain is described by the circuit model shown in Fig.~\ref{fig:A}. Its Hamiltonian can be written as
\begin{equation}
H_{ch} = \frac{1}{2} \sum _{n,m} Q_n C_{nm}^{-1} Q_m + \frac{1}{2L_{SQ}} \left(\frac{\hbar}{2e}\right)^2\sum_n (\phi_n - \phi_{n+1})^2,
\label{eq:A_ham}
\end{equation}
where $Q_n$ and $\phi_n$ denote the charge and the phase of the $n^{th}$ island, respectively. These variables satisfy the canonical commutation relation $[Q_n,\phi_m] = -2ie \delta_{n,m}$. The matrix $C^{-1}_{nm}$ is the inverse of the chain's capacitance matrix
$$
C_{nm} = (C_0 + 2C_{SQ})\delta_{n,m} - C_{SQ} \delta_{n+1,m} - C_{SQ} \delta_{n-1,m}.
$$
We diagonalize the Hamiltonian (\ref{eq:A_ham}) with the help of the following mode expansions for $Q$ and $\phi$
\begin{eqnarray}
\phi_n =  \frac{1}{\sqrt{N}}\sum _k \sqrt{\frac{2 e^2}{C(k) \hbar \omega_k}}(a_k + a^\dagger_{-k})e^{i k n},\label{phiexp}\\
Q_n =  \frac{-ie}{\sqrt{N}}\sum _k \sqrt{\frac{C(k) \hbar \omega_k}{2 e^2}}(a_k - a^\dagger_{-k})e^{i k n}.
\end{eqnarray}
Here, $C(k) = C_0 + 2C_{SQ}(1-\cos k)$ is the discrete Fourier transform of the capacitance matrix, $C(k) = (1/N)\sum_k e^{ik(n-m} C_{nm}$. The dispersion relation is given by
\begin{equation}
\omega_k = \omega_p \sqrt{\frac{2(1 - \cos k)}{C_0/C_{SQ} + 2(1 - \cos k)}},
\end{equation}
with the plasma frequency $\omega_p = 1/\sqrt{L_{SQ} C_{SQ}}$. We use periodic boundary conditions, which implies that $k = 2 \pi m/N$ with $m = \pm 1, 2, \ldots , \pm N/2$. The diagonal form of $H_{ch}$ reads
\begin{equation}
H_{ch} = \sum _k \hbar \omega_k (a^\dagger_k a_k + 1/2).
\end{equation}
The small Josephson junction is connected to one of the ends of the chain, say the one corresponding to $n=0$. As a result, the phase difference $\phi$ across the  the junction acquires a fluctuating part, $\phi_0$, and its Josephson coupling energy can be written as $-E_J \cos (\phi + \phi_0)$. Upon averaging over the fluctuations $\phi_0$, we obtain the junction's effective Josephson energy $U(\phi)$ with a renormalized Josephson coupling energy $E_J^*$,
\begin{equation}
U(\phi) = -E_J \langle \cos (\phi + \phi_0)\rangle_{H_{ch}} = -E_J^*\cos \phi,
\end{equation}
where $E_J^* = E_J \langle \cos \phi_0\rangle_{H_{ch}}$ and the average $\langle \   \rangle_{H_{ch}}$ is taken with respect to the Hamiltonian $H_{ch}$ at zero temperature. Using the fact that $\langle \cos \phi_0\rangle_{H_{ch}} = \exp\{- \langle\phi_0^2\rangle_{H_{ch}}/2\}$ and
the mode expansion (\ref{phiexp}) to calculate $\langle\phi_0^2\rangle_{H_{ch}}$ we finally obtain
\begin{equation}
E_J^*=E_J \exp\left\{-\frac{1}{N} \sum _k \frac{e^2}{\hbar} \sqrt{\frac{L}{2C(k)[1-\cos k]}}\right\}.
\end{equation}
%\begin{figure}
%\includegraphics[angle=0,width=0.5\textwidth]{figure3anew.eps}
%\caption{Dispersion of the propagating modes on Josephson junction chains with a finite capacitance to ground $C_0/C_{SQ} = 1/50$.} \label{fig:6}
%\end{figure}\begin{figure}
%\includegraphics[angle=0,width=0.5\textwidth]{fig3b.eps}
%\caption{Dispersion of the propagating modes on Josephson junction chains with a finite capacitance to ground $C_0/C_{SQ} = 1/50$.} \label{fig:6}
%\end{figure}
\bibliography{Bloch_band_dynamics_v2, revtex-custm}
\end{document}